\documentclass[prl,aps,epsfig,twocolumn]{revtex4}

\usepackage{amsfonts,amssymb,amscd,amsthm}
\usepackage{graphicx}
\usepackage{mathrsfs}
\usepackage[intlimits]{amsmath}
\usepackage[colorlinks, citecolor=red]{hyperref}
\usepackage{physics}
\usepackage{units}
\begin{document}

\title{Probing a dissipative phase transition with a trapped ion through reservoir engineering}

\author{M.-L. Cai\footnote{These authors contribute equally to this work}$^{1}$, Z.-D. Liu$^{*1}$, Y. Jiang$^{*1}$, Y.-K. Wu$^{1}$, Q.-X. Mei$^{1}$, W.-D. Zhao$^{1}$, L. He$^{1}$, X. Zhang$^{2,1}$, Z.-C. Zhou$^{1,3}$, L.-M. Duan\footnote{Corresponding author: lmduan@tsinghua.edu.cn}$^{1}$}
\affiliation{$^{1}$Center for Quantum Information, Institute for Interdisciplinary Information Sciences, Tsinghua University, Beijing 100084, PR China}
\affiliation{$^{2}$Department of Physics, Renmin University, Beijing 100084, PR China}
\affiliation{$^{3}$Beijing Academy of Quantum Information Sciences, Beijing 100193, PR China}

%Dissipation is a ubiquitous phenomenon in physical systems due to their inevitable coupling to the environment. In quantum systems aimed at quantum computing and quantum information tasks, such couplings are necessary for the initialization and the readout of the quantum states, but the resultant dissipation is often regarded as a side effect to be suppressed, which can cause errors in the unitary operations of the system. However, it has been realized that dissipative effects are not always detrimental:
%With suitable dissipation, the steady state of a quantum system can exhibit nontrivial properties and can find applications in engineering many-body entangled states \cite{Diehl2008,PhysRevA.78.042307} and in quantum computing \cite{Verstraete2009}. Furthermore, as one tunes the parameters of a dissipative system, the steady state can undergo abrupt changes owing to the interplay between the coherent drive and the dissipation, which is known as a dissipative phase transition (DPT) \cite{PhysRevA.98.042118}.
\begin{abstract}
Dissipation is often considered as a detrimental effect in quantum systems for unitary quantum operations. However, it has been shown that suitable dissipation can be useful resources both in quantum information and quantum simulation. Here, we propose and experimentally simulate a dissipative phase transition (DPT) model using a single trapped ion with an engineered reservoir. We show that the ion's spatial oscillation mode reaches a steady state after the alternating application of unitary evolution under a quantum Rabi model Hamiltonian and sideband cooling of the oscillator. The average phonon number of the oscillation mode is used as the order parameter to provide evidence for the DPT. Our work highlights the suitability of trapped ions for simulating open quantum systems and shall facilitate further investigations of DPT with various dissipation terms.
\end{abstract}

\maketitle
\textit{Introduction}\textemdash Dissipation is ubiquitous in physical systems, and is often regarded as an undesired error source in quantum information science. However, well-controlled dissipation can also be helpful resources and has found applications in preparing many-body entangled states \cite{Diehl2008,PhysRevA.78.042307}, quantum information processing \cite{Verstraete2009,PhysRevA.83.012304} and the study of nonequilibrium phase transitions \cite{Diehl2008,PhysRevA.86.012116}. In particular, dissipative phase transitions have been observed in various experimental systems such as Bose-Einstein condensate in optical cavities \cite{Brennecke11763,Klinder3290}, semiconductor microcavities \cite{PhysRevLett.118.247402,fink2018signatures} and superconducting circuits \cite{PhysRevX.7.011012,PhysRevX.7.011016,PhysRevLett.122.183601}. However, due to the experimental difficulty in harnessing the dissipation, all these experiments utilize the intrinsic dissipation in the system which can not be tuned. A goal that remains outstanding is to demonstrate a DPT through reservoir engineering to generate a controlled and suitable dissipation.

The trapped ion system makes a desirable platform for studying engineered DPT. As one of the earliest physical systems and a leading one for quantum computing, trapped ions support accurate and coherent manipulation of the quantum states \cite{PhysRevLett.117.060504,PhysRevLett.117.060505}, and can be well isolated from the environment to provide low intrinsic decoherence \cite{RevModPhys.75.281}. Through optical pumping, dissipation in the spin and the motional modes has also been demonstrated to initialize the system \cite{RevModPhys.75.281}, to prepare desired entangled states \cite{Lin2013} and to simulate open system quantum dynamics \cite{Barreiro2011,Schindler2013}. Recently, it has been theoretically proposed that a DPT can be observed using two trapped ions \cite{PhysRevA.97.013825}, with one ion and a collective oscillation mode forming a quantum Rabi model (QRM) \cite{Pedernales2015,PhysRevX.8.021027}, and the second ion being laser cooled to provide a controllable dissipation to the bosonic oscillation mode. Despite being a small system, suitable thermodynamic limit of large number of excitations can be defined as the ratio between the spin and the bosonic mode frequencies increases \cite{PhysRevA.95.012128,PhysRevA.97.013825}, thus allows nonanalytical change across the phase transition point.

\begin{figure}
	\includegraphics[width=1.\linewidth]{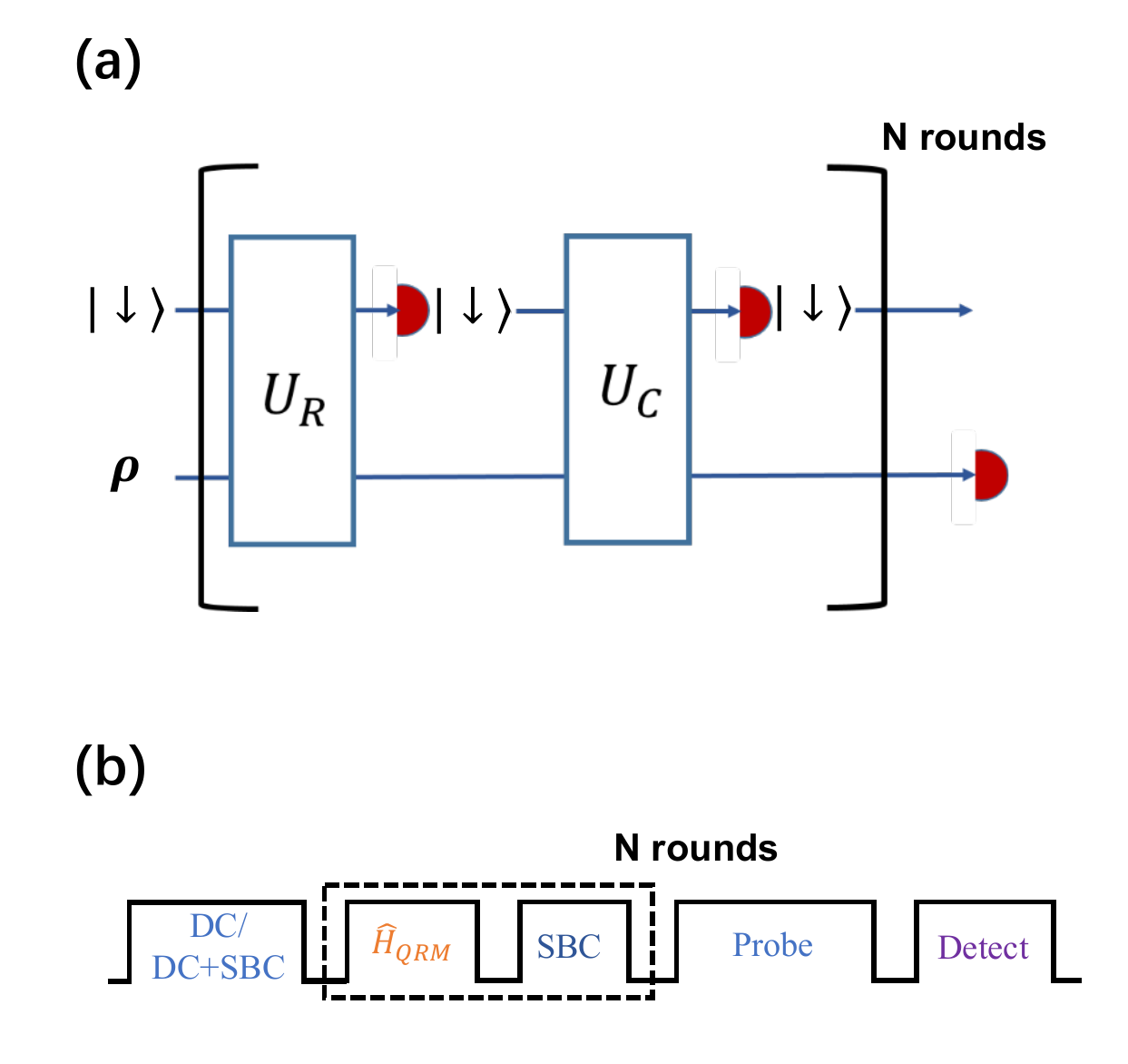}
	\caption{Experimental scheme. (a) An illustration of the experimental sequence in the quantum circuit model. The initial state can be arbitrarily chosen because the final steady state is independent of this choice. Here we initialize the qubit state to $\ket{\downarrow}$ by optical pumping. Then we apply $N$ rounds of alternating coherent drive (unitary evolution $U_R$ under the QRM Hamiltonian) and dissipation (unitary evolution $U_C$ under the red sideband driving sandwiched by two optical pumping stages) to bring the system into a steady state. Finally, we measure the average phonon number $\langle a^\dagger a \rangle$ of the phonon steady state. (b) The complete pulse sequence. Two types of initial phonon states are used: a thermal state after Doppler cooling (DC), or a phonon ground state after additional sideband cooling (SBC). To measure the average phonon number in the final steady state, we apply a probe laser beam on the blue sideband and detect the spin state to fit the phonon population \cite{RevModPhys.75.281} (see SM for details).}
	\label{figure:sequence}
\end{figure}

Here, we propose and experimentally demonstrate a simplified model using only one trapped ion and one oscillator mode, with interleaved pulse sequences of coherent drive and dissipation on the system as shown in Fig.~\ref{figure:sequence}. The system can approach a steady state under these two competing effects and, depending on their relative strength, the steady state can have vanishing phonon number or be strongly driven to high phonon populations to break the $Z_2$ symmetry, thus allows a second-order DPT in the intermediate parameter regime \cite{PhysRevA.97.013825,PhysRevA.98.042118}.
%The steady state is obtained by periodically applying unitary evolution under a QRM Hamiltonian and sideband cooling which creates the engineered dissipation on both the spin and the bosonic mode. We then measure the average phonon number in the steady state, the order parameter for the DPT, near the phase transition point. The change in the average phonon number becomes sharper as we increase the ratio parameter, which indicates the phase transition in the thermodynamic limit.
%

\textit{Experimental scheme}\textemdash In this experiment, the coherent drive is governed by a QRM Hamiltonian \cite{PhysRevX.8.021027},
\begin{equation}
\label{eq:rabi_model}
\hat{H}_{\mathrm{QRM}}=\frac{\omega_a}{2} \hat{\sigma}_{z}+\omega_f \hat{a}^{\dagger} \hat{a}+\lambda\left(\hat{\sigma}_{+}+\hat{\sigma}_{-}\right)\left(\hat{a}+\hat{a}^{\dagger}\right),
\end{equation}
where $\hat{a}^{\dagger}$ ($\hat{a}$) is the bosonic mode creation (annihilation) operator and $\hat{\sigma}_+$ ($\hat{\sigma}_-$) is the spin raising (lowering) operator; $\omega_a$, $\omega_f$ and $\lambda$ are the spin transition frequency, the bosonic mode frequency and the coupling strength between the two subsystems, respectively. This model has been widely studied through quantum simulation in many experimental platforms \cite{PhysRevX.8.021027,Rabi_model,Langford2017,PhysRevLett.108.163601,PhysRevLett.102.186402} including trapped ions \cite{PhysRevX.8.021027,Rabi_model}. In this work, we consider a single $\ensuremath{^{171}\mathrm{Yb}^+~}$ ion in a linear Paul trap (for further details about the setup, see Supplementary Materials (SM)). The spin is encoded in the $\ket{\downarrow}=\ket{{}^{2}S_{1/2},F=0,m_F=0}$ and the $\ket{\uparrow}=\ket{{}^{2}S_{1/2},F=1,m_F=0}$ levels of the ion with atomic transition frequency $\omega_0=2\pi\times 12.6\,\unit{GHz}$, and the bosonic mode is represented by a radial oscillation mode with trap frequency $\omega_{\mathrm{m}}=2\pi\times 2.35\,\unit{MHz}$. We first apply Doppler cooling (DC) to bring the ion into the Lamb-Dicke regime \cite{RevModPhys.75.281}. Then we shine bichromatic Raman laser beams onto the ion to form two pairs of Raman transitions with detuning $\delta_b$ ($\delta_r$) from the blue (red) motional sideband of the oscillation mode. When the two pairs of Raman transitions have the same sideband Rabi frequency $\Omega_{\mathrm{SB}}$, we get an effective QRM Hamiltonian by identifying $\omega_{a}=(\delta_{b}+\delta_{r})/2$, $\omega_{f}=(\delta_{b}-\delta_{r})/2$ and $\lambda=\Omega_{\mathrm{SB}}/2$ in an interaction picture with $\hat{H}_0=(\omega_0-\omega_a)\hat{\sigma}_z/2+(\omega_{\mathrm{m}}-\omega_f)\hat{a}^\dag\hat{a}$ \cite{Rabi_model}. For the calibration of the model parameters $\delta_{b\,(r)}$ and $\Omega_{\mathrm{SB}}$, please see SM.

The intrinsic phonon decoherence rate is estimated to be around $0.2\,\unit{kHz}$ in our system. In order to engineer a strong and controllable dissipation in the system, we employ the sideband cooling process \cite{RevModPhys.75.281} where a laser pulse resonant to the red-sideband transition is sandwiched by two optical pumping stages of the ions to reset the spin state to $\ket{\downarrow}$. In an interaction picture with $\hat{H}_0^\prime=\omega_0\hat{\sigma}_z/2+\omega_{\mathrm{m}}\hat{a}^\dag\hat{a}$, the resonant driving on the red sideband can be represented by $\hat{H}_{c}=\Omega_{c}\left(\hat{a} \hat{\sigma}_{+} +\hat{a}^{\dagger} \hat{\sigma}_{-}\right)/2$, hence after time $\tau_c$, an initial state with $n$ phonons will evolve into $\ket{\downarrow}\ket{n}\to \cos (\sqrt{n}\Omega_c \tau_c/2)\ket{\downarrow}\ket{n} - i\sin (\sqrt{n}\Omega_c \tau_c/2) \ket{\uparrow}\ket{n-1}$. Now after resetting the spin state again through optical pumping, the probability to reduce a phonon is $\sin^2 (\sqrt{n}\Omega_c \tau_c/2)\approx n \Omega_c^2\tau_c^2/4$ assuming $n\lesssim 1/(\Omega_c\tau_c)^2$ (see SM for further discussion when this assumption is broken down), which resembles a phonon damping term with the Lindblad operator $\hat{L}=\Omega_c\sqrt{\tau_c}\hat{a}/2$. Therefore this sideband cooling mechanism can be modeled as a master equation $\dot{\rho}_m=\hat{L}\rho\hat{L}^\dag-\{\hat{L}^\dag\hat{L},\rho_m\} / 2$ where $\rho_m$ is the reduced density matrix in the phonon subspace, together with a reset of the spin state to $\ket{\downarrow}$ after each cycle. This process offers a much stronger dissipation than the intrinsic one for both the spin and the bosonic modes with extraordinary controllability, thus allows us to explore the rich DPT phenomena. Note that small violation of the above approximation condition will slightly decrease the cooling rate for high-phonon-number states, but it shall not change the qualitative behavior of the phase transition.

Through the alternating application of the coherent drive and the artificial dissipation, the system is expected to reach a steady state such that the observables no longer change as we increase the number of cycles. Throughout this work, we consider the average phonon number in the bosonic mode as the order parameter to indicate the phase transition. It can be measured by probing the blue motional sideband (note that at the end of the preceding sideband cooling stage, we have already reset the spin state to $\ket{\downarrow}$) and detecting the spin state to fit the phonon population \cite{RevModPhys.75.281}, as sketched in Fig.~\ref{figure:sequence}(b). Also note that the steady state is expected to be independent of the choice of the initial state. Here we consider two possible initial states, where the phonon state can be either a thermal state generated from Doppler cooling (DC), or the ground state from Doppler cooling followed by sideband cooling (DC+SBC).

\begin{figure}[tbp]
   \includegraphics[width=1.0\linewidth]{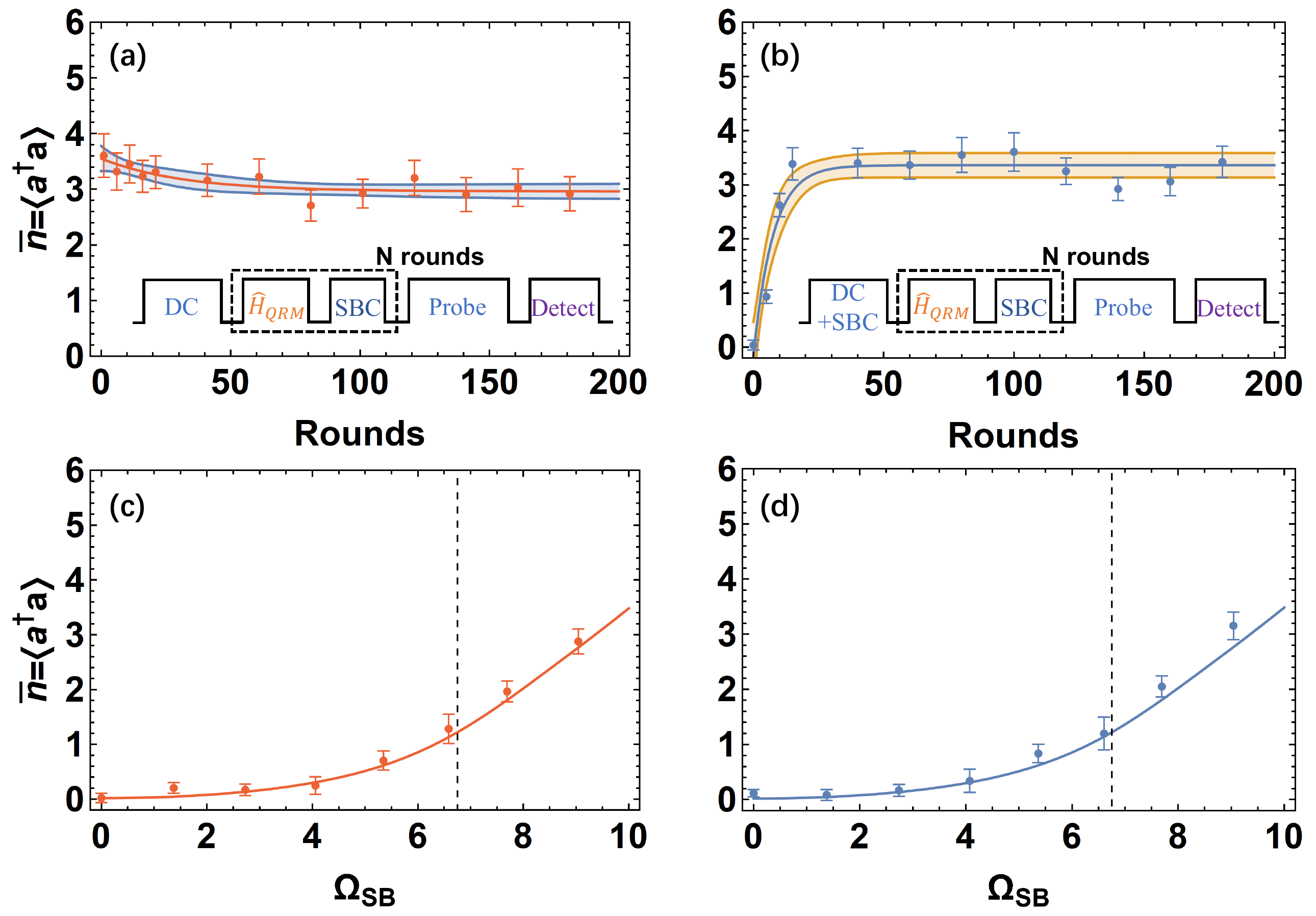}
   \caption{Dynamics and steady state properties at the ratio $R=25$. We set the experimental parameters $\delta_b=2\pi\times26\,\unit{kHz}$ and $\delta_r=2\pi\times24\,\unit{kHz}$, which corresponds to a ratio $R\equiv\omega_{a}/\omega_{f}=(\delta_b+\delta_r)/(\delta_b-\delta_r)=25$. (a) The dynamics of the system approaching the steady state indicated by the average phonon number. The initial phonon state is a thermal state prepared by Doppler cooling. The horizontal axis is the number of rounds of coherent driving and dissipation, as indicated by the inset. In each round, the QRM Hamiltonian is applied for $\tau=20\,\unit{\mu s}$ with a sideband Rabi frequency $\Omega_{\mathrm{SB}}=2\pi\times9.0\,\unit{kHz}$, and the $\tau_d=13\,\unit{\mu s}$ sideband cooling process consists of a $\tau_c=5\,\unit{\mu s}$ driving on the red sideband at $\Omega_c=2\pi\times 20\,\unit{kHz}$ together with the optical pumping and the idle stages. (b) The same plot as (a) for an initial phonon ground state prepared by additional sideband cooling. Note the two "SBC"s in the inset have different meanings, the former "SBC" means a multi-pulse sequence for ground state cooling and the latter "SBC" means a single pulse operation for dissipation. Each dot represents one measured data with the error bar indicating one standard deviation (1 S.D.). The fitting line follows $\bar{n}=Ae^{-N/N_0}+B$ where $A$, $B$ and $N_0$ are fitting parameters, with the shaded region showing a 0.9 confidence level band of the fitting. In both cases, the average phonon number approaches roughly the same value of about $\bar{n}=3.2$ after about $N=50$ cycles, suggesting a steady state independent of the initial state. (c) and (d) The steady state average phonon number for the two initial states as in (a) and (b) respectively versus the sideband Rabi frequency $\Omega_{\mathrm{SB}}$ in the QRM Hamiltonian. The other parameters are unchanged. We fix the number of rounds to 200 to ensure that the final average phonon number has saturated. The colored dots with error bars representing 1 S.D. are the experimental results and the colored lines are from the numerical simulation. The vertical dashed line indicates the numerically-computed phase transition point (see SM for details). A crossover between the two phases can be clearly observed.}
   \label{Fig 0}
\end{figure}

\textit{Experimental results}\textemdash As shown in many previous works (see e.g. Ref.~\cite{PhysRevA.97.013825,Rabi_model}), in such a finite-component system, a thermodynamic limit can be defined as $R\equiv\omega_{a}/\omega_{f}=(\delta_b+\delta_r)/(\delta_b-\delta_r)$ approaches infinity to allow a large number of excitations. We start from $R=25$ in Fig.~\ref{Fig 0} by setting $\delta_b=2\pi\times26\,\unit{kHz}$ and $\delta_r=2\pi\times24\,\unit{kHz}$. We fix the duration of the coherent driving stage in each cycle to be $\tau=20\,\unit{\mu s}$ and vary the sideband Rabi frequency $\Omega_{\mathrm{SB}}$ to study the phase transition. As for the dissipation stage, we drive the red sideband at the sideband Rabi frequency $\Omega_c=2\pi\times 20\,\unit{kHz}$ for $\tau_c=5\,\unit{\mu s}$, which, together with the two $3\,\unit{\mu s}$ optical pumping pulses and the idle time in between, makes up the total $\tau_d=13\,\unit{\mu s}$ duration. In Fig.~\ref{Fig 0}(a) and Fig.~\ref{Fig 0}(b), we present two examples with $\Omega_{\mathrm{SB}} = 2\pi\times9.0\,\unit{kHz}$ for how the average phonon number approaches the steady state value, starting from a thermal state and the ground state of the phonon mode, respectively. Regardless of the initial state, the steady state average phonon number saturates at around $\bar{n}=3.2$, thus verifies that our engineered dissipative term can lead to a unique steady state.
We further fit the data by an exponential decay $\bar{n}=Ae^{-N/N_0}+B$ where $N$ is the number of rounds while $A$, $B$ and $N_0$ are fitting parameters. The fitting results are shown as the central curves with the shaded areas representing a 0.9 confidence level band. It is evident that in these two examples, the system is already reasonably close to the steady state (or at least the value of the average phonon number, which is the relevant observable for the DPT) after about 50 rounds.
The saturation rate also depends on the driving and the cooling parameters, hence in the next step when we scan $\Omega_{\mathrm{SB}}$ to study the change in the average phonon number of the steady state, we increase the number of rounds to 200 to ensure saturation. In Fig.~\ref{Fig 0}(c) and Fig.~\ref{Fig 0}(d) we plot the average phonon number in the steady state versus the sideband Rabi frequency $\Omega_{\mathrm{SB}}$ in the QRM, again for the initial thermal state and the phonon ground state, respectively. The measured steady state phonon numbers match well with the numerical simulations which have already included the decoherence effect of motion (see SM for detailed discussion) and indicate a smooth crossover between the two phases. The vertical dashed line indicates the phase transition point from the numerical calculation (see SM for details). For a real transition from the normal phase to the superradiance phase \cite{PhysRevA.97.013825} across this line, we expect that the average phonon number would have a nonanalytical increasement when the ratio $R$ approaches infinity. This is similar to the quantum phase transition in the closed QRM \cite{Rabi_model} where the nonanalytical behavior happens on the ground state rather than the steady state. However, due to the finite ratio $R=25$ we adopted here, we can only see a smooth crossover behavior.

\begin{figure}[tbp]
   \includegraphics[width=0.95\linewidth]{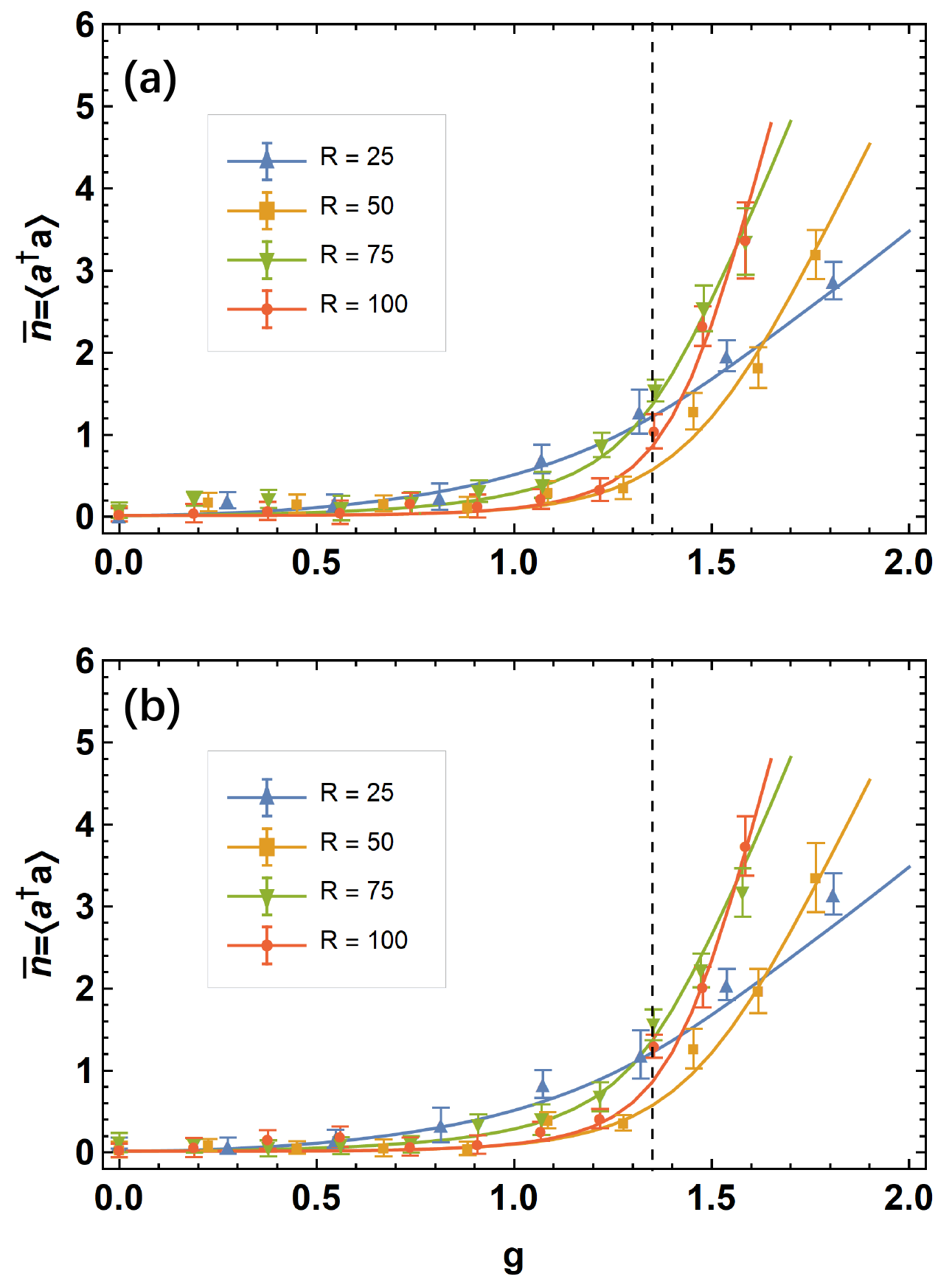}
   \caption{Average phonon number under the increasing ratio $R$. Here we repeat the measurement for the steady state average phonon number in Fig.~2 for different ratio parameters $R=50,\,75,\,100$ by keeping $(\delta_b-\delta_r)/2\pi=2\,\unit{kHz}$ fixed while increasing $(\delta_b+\delta_r)/2\pi$ from $50\,\unit{kHz}$ to $100\,\unit{kHz}$, $150\,\unit{kHz}$ and $200\,\unit{kHz}$. Under each $R$, we measure the average phonon number in the steady state versus the sideband Rabi frequency $\Omega_{\mathrm{SB}}$ starting from (a) a thermal state or (b) the phonon ground state similar to Fig.~2. The horizontal axis is the dimensionless coupling $g\equiv 2\Omega_{\mathrm{SB}}/\sqrt{\delta_b^2-\delta_r^2}$, and the vertical dashed line indicates the numerically-computed phase transition point at $g_c \approx 1.35$. The colored dots are experimental data with error bars representing 1 S.D.. The colored curves are the results from numerical simulations under the same parameters. The measured data agree with the numerical results within about 1 S.D., and we can see that the sharpness of the curve increases with the ratio parameter. The similarity between (a) and (b) again verifies that the steady states are independent of the initial states.}
   \label{Fig 1}
\end{figure}

To acquire further evidence of the DPT, we study the finite frequency scaling behavior under the increasing ratio $R$, which corresponds to the system size in the conventional thermodynamic limit. Here we fix $\delta_b-\delta_r=2\pi\times 2\,\unit{kHz}$ and increase $\delta_b+\delta_r$ up to $2\pi\times 200\,\unit{kHz}$ and study the scaling behavior of the average phonon number, as shown in Fig.~\ref{Fig 1}. Again we consider two different initial states and find that the steady state properties are unaffected. The experimental results agree well with the numerical prediction within about 1 S.D.. The main error source of the deviation can be referred to the SM. Again, the numerical simulation results have already considered the decoherence effect of motion. As we can see, the change in the average phonon number becomes sharper with increasing $R$. However, the phonon number shows a nonmonotonic behavior with increasing $R$ near the critical point, this is further invesgated in the SM. Besides, the numerical simulation shows that for $g$ below the critical point $g_c$, the average phonon number converges to a finite values as we increase $R$, while for $g>g_c$, the steady state phonon number diverges in the limit $R \to \infty$. These behaviors indicate a DPT in the thermodynamic limit. More numerical and experimental results are presented in SM to prove the existence of phase transition in this model.

\begin{figure}[tbp]
   \includegraphics[width=0.95\linewidth]{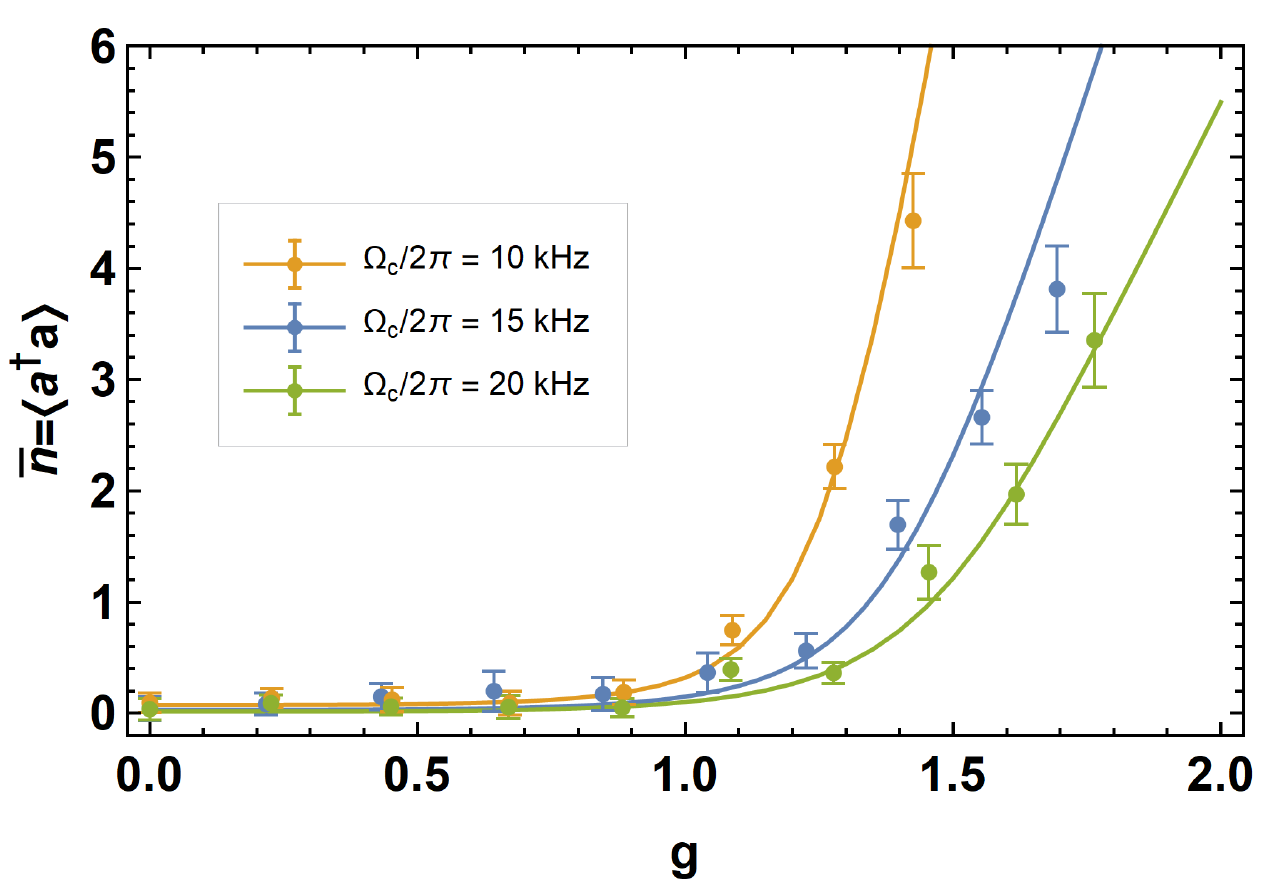}
   \caption{Average phonon number under different cooling rates. Here we fix the ratio $R=50$ and tune the cooling rate by varying the red sideband Rabi frequency $\Omega_c$. With different $\Omega_c$, we measure the change of the average steady-state phonon number versus the dimensionless coupling strength $g$. The colored dots are experimental data with error bars representing 1 S.D.. The colored curves are the results from numerical simulations under the same parameters. The measured data agree with the numerical results within about 1 S.D., and we can see that as the cooling rate increases, the transition point (crossover region) is shifted to higher $g$ due to the competing effect between the driving and the dissipation terms.}
   \label{Fig 2}
\end{figure}

Finally we look into the influence of the cooling rate, which is enabled by our engineered reservoir, at the fixed frequency ratio $R=50$. With other parameters unchanged, the cooling rate can be easily tuned by varying the red sideband Rabi frequency $\Omega_\mathrm{c}$. Here we start from the phonon ground state for simplicity since the steady state properties have been verified above to be independent of the initial state. For various $\Omega_\mathrm{c}$ from $2\pi \times 10\,\unit{kHz}$ to $2\pi \times 20\,\unit{kHz}$, we measure the average steady-state phonon number versus the dimensionless coupling strength $g\equiv 2\Omega_{\mathrm{SB}}/\sqrt{\delta_b^2-\delta_r^2}$, with the experimental results shown in Fig.~\ref{Fig 2}. Clearly, with larger cooling rates, the transition point (crossover region for finite $R$) moves toward higher $g$ due to the competition between the coherent driving and the dissipation terms.

\textit{Discussion and conclusions}\textemdash To sum up, we propose and demonstrate a DPT model with an artificially engineered reservoir using a single trapped ion. We first verify the steady states of the ion motion and clearly observe a crossover between two different phases of the phonon mode. Second, we implement the finite frequency scaling to study the DPT where the crossover becomes sharper with an increasing frequency ratio. Finally, we observe the behavior of the crossover under different dissipative rates via tuning the sideband cooling rate. To our knowledge, our experiment is the first experimental probe of a DPT through reservoir engineering with adjustable dissipation rates. It shows the advantage of strong controllability of the trapped-ion system for the simulation of open quantum systems and shall facilitate further investigations of DPT under various engineered dissipation terms. Also, the demonstrated scheme here is a universal method very similar to those used in dissipative nonclassical-state engineering \cite{Kienzler53} and the pulsed-CPT scheme in quantum metrology \cite{PhysRevLett.116.043603,Nicolas_2018}. In this sense, the scheme can be well adapted to other research fields where a controlled dissipation is desired. The observation of a critical phenomenon near the transition point and experimentally extracting the critical exponent are basically limited by the frequency ratio we can achieve under the current experimental conditions. With the suppression of the fluctuation of the experimental parameters ($\delta_b$, $\delta_r$), we can further decrease the denominator of the ratio ($\delta_b-\delta_r$) without too much deviation, hence an observation of the critical phenomena becomes possible.

This work was supported by the Beijing Academy of Quantum Information Sciences, the National key Research and Development Program of China (2016YFA0301902), Frontier Science Center for Quantum Information of the Ministry of Education of China, and Tsinghua University Initiative Scientific Research Program. Y.-K. W. acknowledges support from Shuimu Tsinghua Scholar Program and International Postdoctoral Exchange Fellowship Program (Talent-Introduction Program).

\bigskip

%\bibliographystyle{ieeetr}{}
%\bibliographystyle{unsrt}{}
%\bibliographystyle{apsrev4-1}
%%%%%%%%%%%%%%%%%%%%%%%%%%%%%%%%%%%%%%%%{}{}

%%%%%%%%%%%%%%%%%%%%%%%%%%%%%%%%%%%%%%
\end{document}

\title{Supplementary Materials for ``Probing a dissipative phase transition with a trapped ion through reservoir engineering''}

\author{M.-L. Cai\footnote{These authors contribute equally to this work}$^{1}$, Z.-D. Liu$^{*1}$, Y. Jiang$^{*1}$, Y.-K. Wu$^{1}$, Q.-X. Mei$^{1}$, W.-D. Zhao$^{1}$, L. He$^{1}$, X. Zhang$^{2,1}$, Z.-C. Zhou$^{1,3}$, L.-M. Duan\footnote{Corresponding author: lmduan@tsinghua.edu.cn}$^{1}$}
\affiliation{$^{1}$Center for Quantum Information, Institute for Interdisciplinary Information Sciences, Tsinghua University, Beijing 100084, PR China}
\affiliation{$^{2}$Department of Physics, Renmin University, Beijing 100084, PR China}
\affiliation{$^{3}$Beijing Academy of Quantum Information Sciences, Beijing 100193, PR China}

\maketitle
\section{Experimental setup}

Our experimental setup is a single $\ensuremath{^{171}\mathrm{Yb}^+~}$ ion in a linear Paul trap. The spin state is encoded in the $\ket{\downarrow}=\ket{{}^{2}S_{1/2},F=0,m_F=0}$ and the $\ket{\uparrow}=\ket{{}^{2}S_{1/2},F=1,m_F=0}$ levels of the ion with atomic transition frequency $\omega_0=2\pi\times 12.6428\,\unit{GHz}$, and the bosonic mode is encoded in a radial oscillation mode with secular frequency $\omega_{\mathrm{m}}=2\pi\times 2.35\,\unit{MHz}$.

We use counter-propagating $355\,\unit{nm}$ pulsed laser beams with a repetition rate $\omega_{\mathrm{rep}}\approx 2\pi \times 118.415\,\unit{MHz}$ and a bandwidth of about $200\,\unit{GHz}$ to manipulate the qubit through Raman transition. Two acousto-optic modulators (AOMs) are used to fine-tune the frequency and the amplitude of the Raman transitions. More details can be found in Ref.~[24] of the main text.

\section{Measurement of phonon population}

We follow the standard method of Ref.~[15,24] of the main text to fit the phonon state population. Note that after each cycle of coherent drive and dissipation, the spin state is already pumped to $\ket{\downarrow}$, so we only need to apply a blue sideband pulse with various duration $t$ and measure the spin-up state population $P_{\uparrow}$ afterwards. It can be fitted by
\begin{equation}
P_{\uparrow}(t)=\frac{1}{2}\left[1 - \sum_{k=0}^{k_{\mathrm{max}}} p_k e^{-\gamma_k t}\cos (\Omega_{k,k+1} t)\right],
\end{equation}
where $p_k$ is the occupation in the Fock state $\ket{k}$, $\gamma_k$ a number-state-dependent empirical decay rate of the Rabi oscillation, $\Omega_{k,k+1}\propto\sqrt{k+1}$ the number-state-dependent sideband Rabi frequency, and $k_{\mathrm{max}}$ the cutoff in the phonon number. After fitting the phonon state population $P=(p_0,\,p_1,\,\cdots)^T$ with its covariance matrix $\Sigma$, we can compute the average phonon number $\bar{n} = N \cdot P$ where $N=(0,\,1,\,\cdots)$ is a row vector representing the phonon number basis. Assuming the fitted parameters follow a joint Gaussian distribution, we can estimate the variance of $\bar{n}$ as $\sigma_{\bar{n}}^2 = N \Sigma N^T$, hence the error bar of the average phonon number of one standard deviation is $\sigma_{\bar{n}}=\sqrt{ N \Sigma N^T }$.

\section{The calibration of model parameters}
In the quantum Rabi model Hamiltonian (see formula (1) in the main text), three parameters $\omega_{a}$, $\omega_{f}$ and $\lambda$ fully determine the Hamiltonian. In experiment simulation with trapped ion, $\omega_{a}=(\delta_b+\delta_r)/2$, $\omega_{f}=(\delta_b-\delta_r)/2$ and $\lambda=\Omega_{\mathrm{SB}}/2$ where $\delta_{b\,(r)}$ is the detuning of the differential Raman laser frequency from the blue (red) sideband of the motional mode and $\Omega_{\mathrm{SB}}$ is the sideband Rabi frequency. Before every experiment, we need to calibrate the actual value of $\delta_b$, $\delta_r$ and $\Omega_{\mathrm{SB}}$. In terms of the calibration of $\delta_b$ and $\delta_r$, we first use the microwave Ramsey spectroscopy to determine the qubit frequency $\omega_{\mathrm{q}}$ whose measurement precision is on the order of $2\pi\times5\,\unit{Hz}$; Then we use the Raman sideband Ramsey spectroscopy to determine the secular frequency $\omega_{\mathrm{m}}$ of the ion motion whose measurement precision is on the order of $2\pi\times100\,\unit{Hz}$; Finally, we set the two differential Raman laser frequencies at $\omega_{\mathrm{q}}+\omega_{\mathrm{m}}+\delta_b$ and $\omega_{\mathrm{q}}-\omega_{\mathrm{m}}+\delta_r$ respectively. Using the above method, the deviation of the actual value of $\delta_{b\,(r)}$ from the target value of $\delta_{b\,(r)}$ can be well bounded by the measurement precision of the secular frequency, i.e. the order of $2\pi\times100\,\unit{Hz}$. This means the ratio parameter uncertainty for $R=25$, $50$, $75$ and $100$ is $\pm 1.3$, $\pm 2.5$, $\pm 3.8$ and $\pm 5.0$ respectively, i.e. roughly $5\%$ relative error. Also the trap frequency fluctuation during the measurement is on the order of $2\pi \times 200\, \unit{Hz}$ (considering the $200\,\mathrm{s}^{-1}$ motional dephasing rate described below), this will induce about $10\%$ ratio fluctuation in the experiment. This ratio calibration uncertainty and the ratio fluctuation during the measurement are the main error sources of the experimental data. In terms of the calibration for the sideband Rabi frequency $\Omega_{\mathrm{SB}}$, we just fix the laser beam intensity and scan the sideband Rabi oscillation for several cycles and fit out the Rabi frequency. The laser intensity fluctuation is below $1\,\%$, hence the fluctuation of Rabi frequency is in the same order. This error has tiny effect to experiment data.

\section{Note on the breakdown of the linear approximation of the red sideband pulse}
The validity of our experiment is based on the assumption that we need to make sure that the cooling time duration $\tau_c$ is much smaller than the inverse of the red sideband Rabi rate $\Omega_{c}$, therefore $\sin^2 (\sqrt{n}\Omega_c \tau_c/2)$ can be linearized if the phonon number $n$ is not too large. The consequence of going beyond is that when the experiment steps into the regime where the average phonon number of the state is too large, the terms in the high phonon number cannot be efficiently cooled down by using the red sideband pulse plus the optical pumping. Therefore, the steady state is hard to be reached in a practical numerical simulation time and may be very different from the prediction of the current model. However, with the average phonon number below 10, we have verified that the numerical simulations with or without the linear approximation have nearly the same consequence. Therefore, in our experimental regime, the red sideband pulse plus the spin reset can be well approximated by the dissipative channel $\hat{L}=\Omega_c\sqrt{\tau_c}\hat{a}/2$.

\section{Numerical and experimental investigation of the dissipative phase transition}

\subsection{Numerical investigation}
To further prove the existence of a dissipative phase transition in our model and to understand the critical behavior, here we present numerical results for the finite-size scaling as the frequency ratio $R\equiv(\delta_b+\delta_r)/(\delta_b-\delta_r)$ approaches infinity. To obtain the steady state, we alternatingly simulate the unitary evolution under $\hat{H}_{\mathrm{QRM}}$ and the dissipative process governed by the Lindblad superoperator $\hat{L}=\Omega_c\sqrt{\tau_c}\hat{a}/2$ together with a spin reset as described in the main text. Note that these two processes are described in two different interaction pictures with $\Delta \hat{H}_0 \equiv \hat{H}_0'-\hat{H}_0=\omega_a\hat{\sigma}_z/2+\omega_f\hat{a}^\dag\hat{a}$, thus a time-dependent relative phase needs to be included in the simulation when switching between the two interaction pictures. We repeat these two processes until the calculated average phonon number $\langle \hat{a}^\dag \hat{a} \rangle$ converges.

\begin{figure*}[tbp]
	\includegraphics[width=0.8\linewidth]{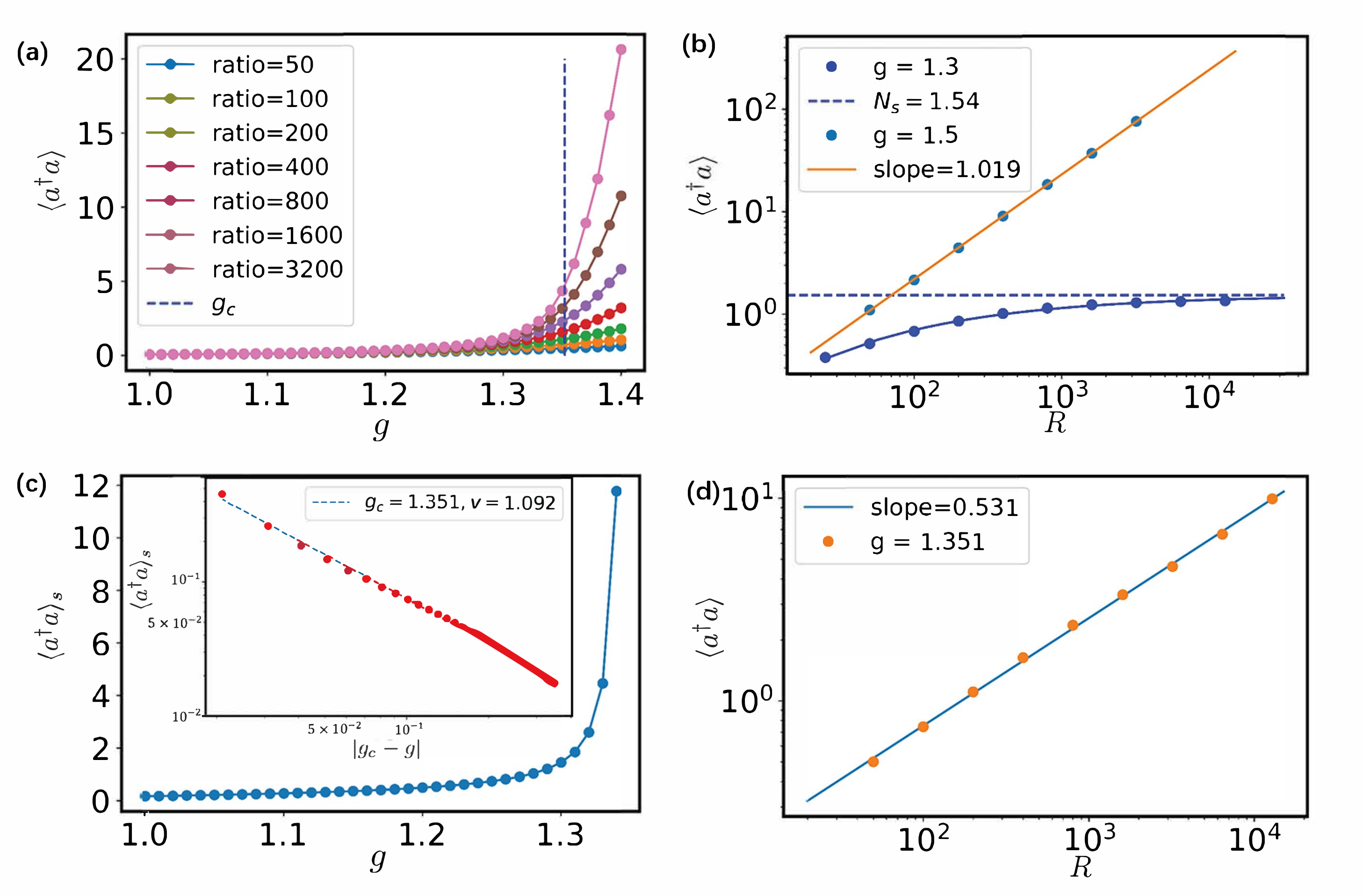}
	\caption{\textbf{Numerical study of the dissipative phase transition.} \textbf{(a)} The steady state phonon number $\langle \hat{a}^\dag \hat{a} \rangle$ versus the dimensionless coupling strength $g$ at various ratio $R$ ranging from $50$ to $3200$. \textbf{(b)} The steady state phonon number  $\langle \hat{a}^\dag \hat{a} \rangle$ versus the ratio parameter $R$ at different coupling strengths ($g=1.3$ and $g=1.5$). In the weak coupling region ($g = 1.3$) the phonon number saturates as $R\to \infty$, while in in the strong coupling region ($g=1.5$), the phonon number shows a power-law scaling with $R$ and approaches infinity as $R\to \infty$. \textbf{(c)} The saturation value of the steady state phonon number $N_s\equiv \langle \hat{a}^\dag \hat{a} \rangle_{s}$ in the limit $R\to\infty$ in the region $g<g_c$. By fitting $N_s = C(g_c-g)^{-\nu}$, we get the critical point $g_c \simeq 1.351$ and a critical exponent $\nu \simeq 1.092$, the error bar is 1 S.D. from fitting. \textbf{(d)} The scaling behavior near the numerically calculated critical point $g_c = 1.351$, which fits a slope of $0.531$ in the log-log plot.}
	\label{fig:numerical}
\end{figure*}
Here we consider the same parameters as in the main text. That is, in each cycle the quantum Rabi model is applied for $\tau=20\,\unit{\mu s}$ at varying sideband Rabi frequency $\Omega_{\mathrm{SB}}$; the sideband cooling pulse with $\Omega_c=2\pi\times20\,\unit{kHz}$ is applied for $\tau_c = 5\,\unit{\mu s}$; additional $10\,\unit{\mu s}$ for optical pumping and idling time are included in the sideband cooling stage to give a total duration $\tau_d=15\,\unit{\mu s}$ when computing the relative phase mentioned above. We fix $\delta_{b}-\delta_{r} = 2\pi\times 2\,\unit{kHz}$ and vary $\delta_{b}+\delta_{r}$ to set the ratio $R$ from $50$ to $3200$. As shown in Fig.~\ref{fig:numerical}(a), we plot the average phonon number $\langle \hat{a}^\dagger \hat{a} \rangle$ versus the dimensionless coupling $g\equiv 2\Omega_{\mathrm{SB}}/\sqrt{\delta_b^2-\delta_r^2}$ from $g=1.0$ to $g=1.4$. For $g$ below $g_c\approx 1.351$, the average phonon number in the steady state saturates to a finite value as we increase $R$; while for $g>g_c$, the steady state phonon number diverges in the limit $R\to\infty$ (note that for increasing $R$ we need to use larger phonon number cutoff in the numerical simulation to suppress the truncation error). In Fig.~\ref{fig:numerical}(b) we plot the scaling behavior of the steady-state phonon number with respect to the ratio $R$ at fixed $g$, and fit the numerical results to obtain the asymptotic form as $R$ approaches infinity. As shown in the figure, for $g=1.3$ in the weak coupling region the phonon number eventually saturates at $N_s= 1.54$, while for $g=1.5$ in the strong coupling region a power-law scaling indicates an infinite steady-state phonon number as $R\to\infty$. This phenomenon of a diverging phonon number is similar to the case considered in Ref.~[22] in the main text. In Fig.~\ref{fig:numerical}(c) we plot the saturation value $N_s$ of the steady state phonon number in the limit $R\to\infty$ versus $g$ in the $g<g_c$ region. We further fit the data near the critical point as $N_s = C (g_c-g)^{-\nu}$ as shown in the inset, from which we extract the critical point $g_c \simeq 1.351 \pm 0.002$ and a critical exponent $\nu \simeq 1.092 \pm 0.029$, the error bar is 1 S.D. from fitting. In Fig.~\ref{fig:numerical}(d), we plot the scaling behavior of the phonon number at the critical point $g_c=1.351$, which fits a power-law behavior $\langle a^\dagger a \rangle \propto R^{0.531}$.

\subsection{Experimental investigation}
\begin{figure*}[tbp]
  \includegraphics[width=0.5\linewidth]{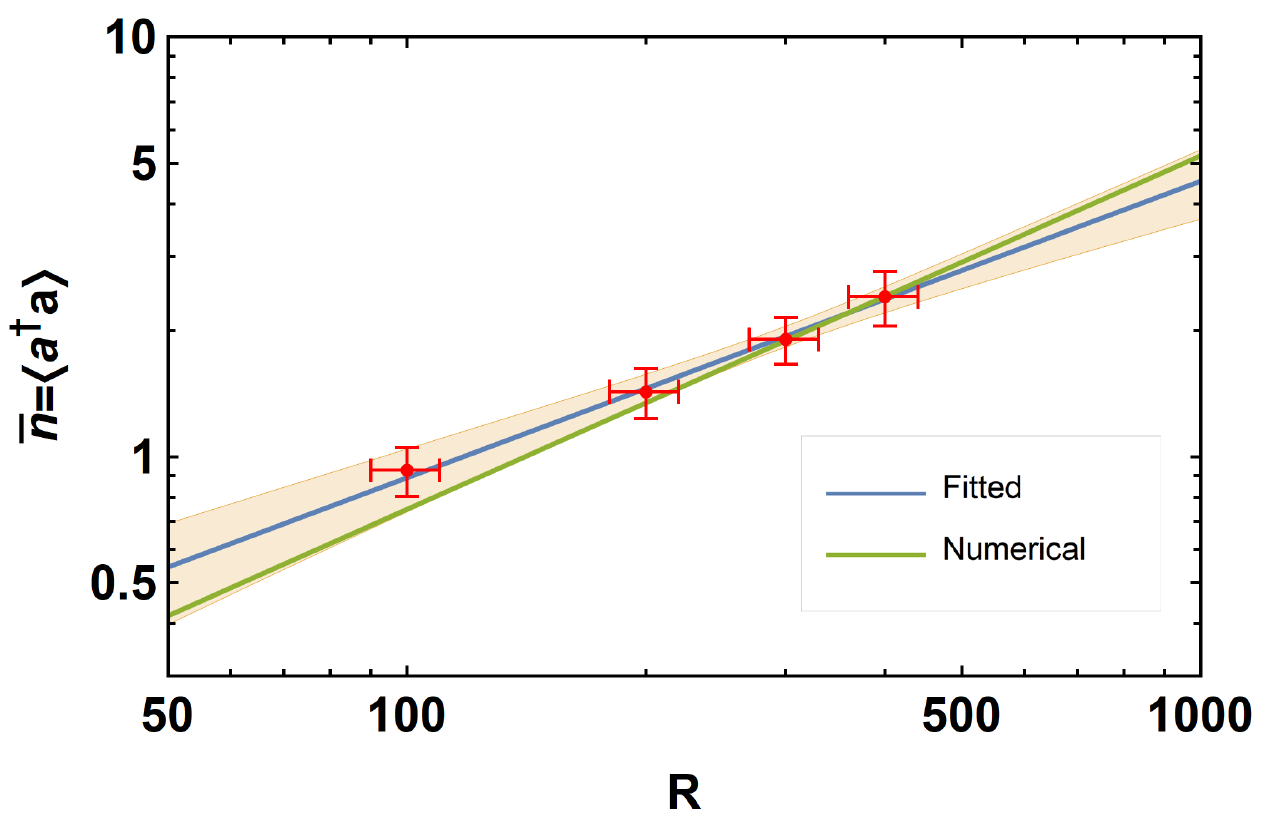}
  \caption{\textbf{Average phonon number scaling.} The red dots are exprimental results with vertical error bars representing 1 S.D. and horizontal error bars estimated from calibration uncertainty. The blue line is a fitting result with the orange shaded region representing 0.95 confidence level (2 S.D.) band. The green line is from numerical simulation whose slope is around $0.843$ representing the critical exponent.}
  \label{fig:scaling}
\end{figure*}

In order to see clear phonon number scaling with finite frequency ratio, we fix the coupling strength $g=1.5$ and try to observe the average phonon number at different frequency ratio just like the result in Fig.~\ref{fig:numerical}(b). In order to have sufficiently large ratio $R$ and experimentally feasible sideband Rabi rate $\Omega_{\mathrm{SB}}$ (in our current condition, $\Omega_{\mathrm{SB}}$ need to be smaller than $2\pi\times20\,\unit{kHz}$ due to the laser intensity limitation) at $g=1.5$, the $\delta_{b}-\delta_{r}$ need to be as small as possible. On the other hand, the precision of the parameter calibration is around $100Hz$. Thus we fix $\delta_{b}-\delta_{r}=1\,\unit{kHz}$ as a tradeoff, resulting in the required Rabi rate $\Omega_{\mathrm{SB}}=2\pi\times7.5,\,10.6,\,13.0,\,15.0 \,\unit{kHz}$ at $R=100,\,200,\,300,\,400$, respectively with around $10\,\%$ relative erorr in $R$. The result is shown in Fig.~\ref{fig:scaling}. We fit the experimental data with a linear line under the log-log scale shown as a blue line in the figure. The orange shaded region represents 0.95 confidence level (2 S.D.) band and the extracted slope of the blue line is $0.707\pm0.148\,(2\,\mathrm{S.D.})$. The green line is calculated from the numerical simulation and the slope is around $0.843$. We can see the experimentally extracted slope is smaller than the numerical result but is still reasonable considering relatively large error of the ratios. A more precise extraction of the critical exponent is limited by the current experimental noise. We need to further reduce the fluctuation of the experimental parameters such as the trap frequency, laser intensity etc. towards an observation of critical phenomena under this model.

\section{Numerical simulation with small ratio increment}
\begin{figure*}[tbp]
	\includegraphics[width=0.8\linewidth]{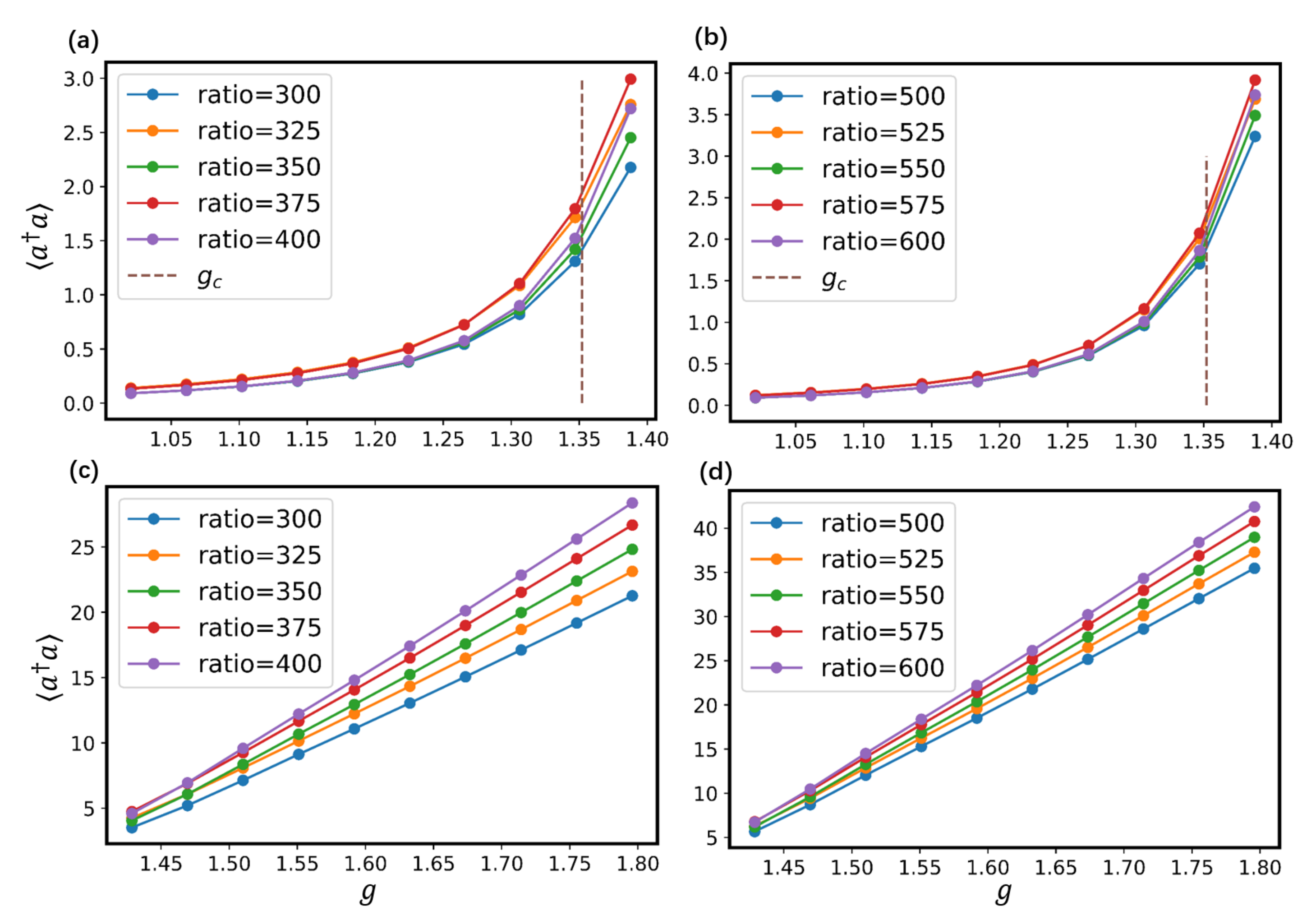}
	\caption{\textbf{The trend with small ratio increment.} We plot the steady state phonon number with respect to different dimensionless coupling $g$ with a small ratio increment $25$ from $300$ to $400$ in \textbf{(a)}, \textbf{(c)} and $500$ to $600$ in \textbf{(b)}, \textbf{(d)}. \textbf{(a)} and \textbf{(b)} show the trend in the near-critical-point regime. \textbf{(c)} and \textbf{(d)} show the trend far beyond the critical point.}
	\label{fig:trend}
\end{figure*}

As the ratio increment is large enough, as shown in Fig.~\ref{fig:numerical} the trend of the phonon number near the critical point shows a good phase transition behavior. However, in this dissipative phase transition model, we find that as the increment of ratio is small, the phonon number of the steady state has a small "back and forth" behavior, and shows a nonmonotonic behavior. In our simulation, we take two different series of the ratios, i.e. ($300$, $325$, $350$, $375$, $400$) and ($500$, $525$, $550$, $575$, $600$). We plot the steady state phonon number versus the dimensionless coupling coefficient $g$ in two different regimes. The first regime is near the critical point, where $g$s are taken from $1.0$ to $1.4$. In this regime, as we increase the ratio with a small value ($25$), the steady state phonon number exhibits a small "back and forth" behavior as shown in Fig. ~\ref{fig:trend} (a), (b). As the dimensionless coupling coefficient is far beyond the critical point (the second observation regime), the trend of the phonon number becomes monotonic, and no "back and forth" behavior is observed. Hence, although a small increment in the ratio may lead to a small "back and forth" behavior in the phonon number, the overall trend of our model still shows a clear phase transition.

\section{Numerical simulation with noise}

We further consider the decoherence effect in the numerical simulation. This can be simulated by invoking the Lindblad superoperator $L[\hat{O}]\hat{\rho} \equiv \hat{O} \hat{\rho} \hat{O}^\dag - \hat{O}^\dag \hat{O}\hat{\rho} / 2 - \hat{\rho} \hat{O}^\dag \hat{O} / 2$. For motional heating and dephasing, the superoperator is $L[\sqrt{\gamma n_{\mathrm{th}}}\hat{a}^\dag]+L[\sqrt{\gamma (n_{\mathrm{th}}+1)}\hat{a}]$ and $L[\sqrt{2\Gamma_m}\hat{a}^\dag\hat{a}]$ respectively \cite{PhysRevA.62.053807}, where $\gamma n_{\mathrm{th}}$ is the heating rate and $\Gamma_{\mathrm{m}}$ is the dephasing rate. In our measurement, the motional dephasing rate $\Gamma_\mathrm{m}$ is around $200\,\mathrm{s}^{-1}$, the heating rate is below $50\,\mathrm{s}^{-1}$. It is not necessary to consider the effect of the spin dephasing because the duration ($\approx 20\,\unit{\mu s}$) between the two spin resets is much smaller than the spin dephasing time of our system ($\approx 50\,\unit{ms}$). 

Besides, we consider another heating effect caused by the photon recoil from the optical pumping. Note that only when the ion is in the spin up state (i.e. the bright state) will it absorb photons. Hence the number of photons being absorbed by an ion is $N_b = N_p \times P_{\uparrow}$, where $N_p$ indicates the average number of photons being absorbed by an ion during the optical pumping (in $\ensuremath{^{171}\mathrm{Yb}^+~}$, $N_p = 3$), and $P_{\uparrow}$ is the population of the spin up state. The heating energy of the ion for each pumping step roughly equals to the recoil energy of the photons, which increases one of the three motional modes phonon number by $\frac{(N_b \hbar k)^2}{2 m \times 3 \omega_{\mathrm{m}}}$ with the photon wavevector $k$, the ion mass $m$ and the motional mode frequency $\omega_{\mathrm{m}}$.

\begin{figure*}[tbp]
  \includegraphics[width=0.5\linewidth]{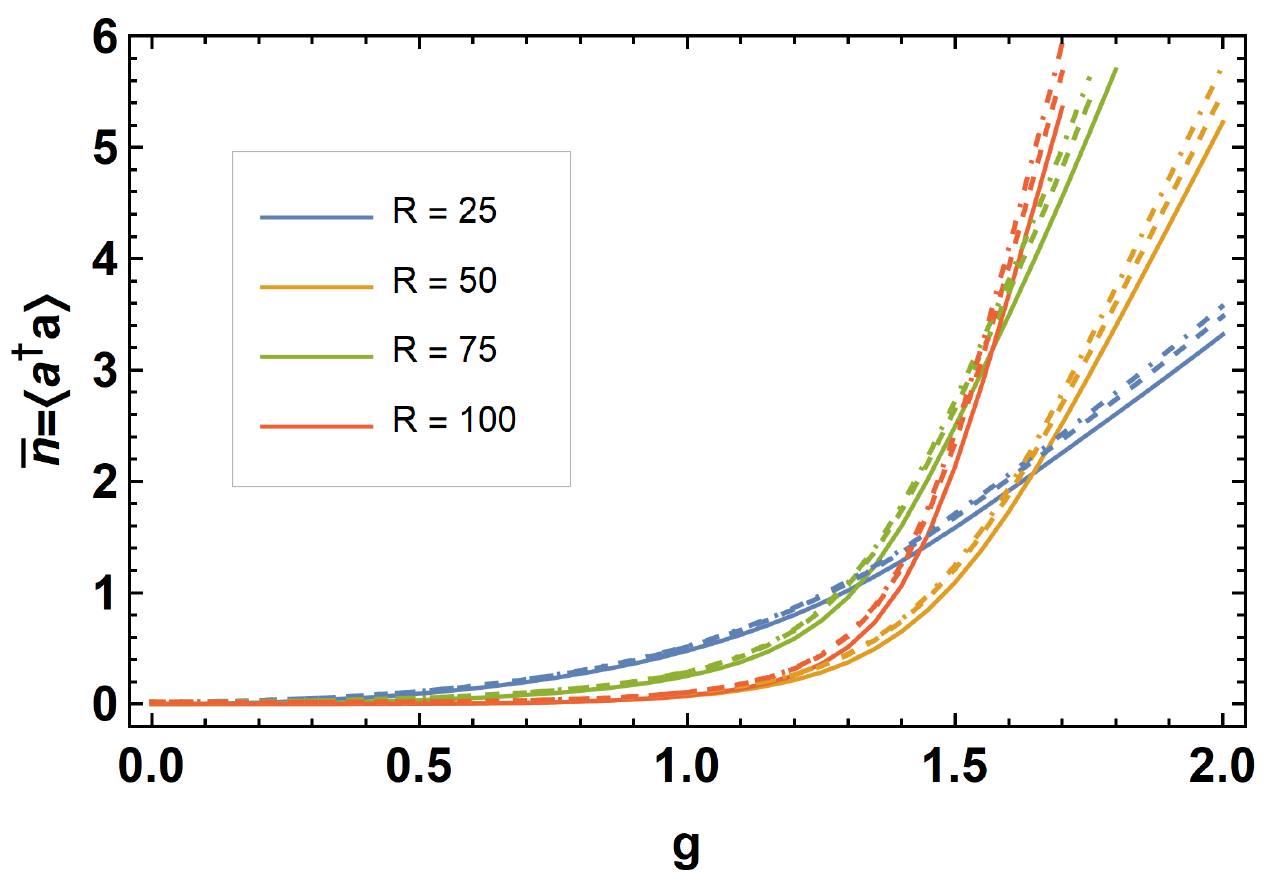}
  \caption{\textbf{Numerical simulation with/without noise effect.} The solid, dashed and dot-dashed lines are the simulation results of the variation of average phonon number versus the coupling strength $g$, without any noise effect, with only the decoherence effect and with both the decoherence and recoil effect, respectively. It's obvious that the phonon number is slightly larger after considering the decoherence effect and recoil effect.}
  \label{fig:decoherence}
\end{figure*}

As shown in Fig.~\ref{fig:decoherence}, the solid, dashed and dot-dashed lines are the simulation results of the variation of average phonon number versus the coupling strenghth $g$, without any noise effect, with only the decoherence effect and with both the decoherence and recoil effect, respectively. We can see the phonon number is slightly larger after considering the decoherence and recoil effect which is consistent with intuition.